\documentclass{article}

\usepackage[top=1.0in, bottom=1.25in, left=0.875in, right=0.875in]{geometry}

\usepackage[cmex10]{amsmath}

\usepackage{amssymb}
\usepackage{graphicx}
\usepackage{float}
\usepackage{tabularx}
\usepackage{multirow}
\usepackage{authblk}  
\usepackage{subcaption}
\usepackage{rotating}  

\let\oldtabular\tabular
\renewcommand{\tabular}{\footnotesize\oldtabular}

\begin{document}
\title{Non-contact hemodynamic imaging reveals the jugular venous pulse waveform}
\date{}

\author[1,2,*]{Robert Amelard}
\author[2,3]{Richard L Hughson}
\author[2]{Danielle K Greaves}
\author[2]{Kaylen J Pfisterer}
\author[1]{Jason Leung}
\author[1]{David A Clausi}
\author[1,2]{Alexander Wong}
\affil[1]{Systems Design Engineering, University of Waterloo, Waterloo, Canada, N2L3G1}
\affil[2]{Schlegel-University of Waterloo Research Institute for Aging, Waterloo, Canada, N2J0E2}
\affil[3]{Applied Health Sciences, University of Waterloo, Waterloo, Canada, N2L3G1}

\maketitle

%
%
\begin{abstract}
Cardiovascular monitoring is important to prevent diseases from progressing. The jugular venous pulse (JVP) waveform offers important clinical information about cardiac health, but is not routinely examined due to its invasive catheterisation procedure. Here, we demonstrate for the first time that the JVP can be consistently observed in a non-contact manner using a novel light-based photoplethysmographic imaging system, coded hemodynamic imaging (CHI). While traditional monitoring methods measure the JVP at a single location, CHI's wide-field imaging capabilities were able to observe the jugular venous pulse's spatial flow profile for the first time. The important inflection points in the JVP were observed, meaning that cardiac abnormalities can be assessed through JVP distortions. CHI provides a new way to assess cardiac health through non-contact light-based JVP monitoring, and can be used in non-surgical environments for cardiac assessment.

\end{abstract}

\flushbottom
%
%

%
%
\section*{Introduction}
Cardiovascular disease is the leading cause of mortality, resulting in 17.3~million deaths per year globally, and a third of all deaths in the United States~\cite{ama2016}. Cardiovascular monitoring is essential to assessing and maintaining or enhancing quality of life through preventive and acute care. The jugular venous pulse (JVP) waveform is a powerful diagnostic tool for assessing cardiac function. The jugular vein is a major venous extension of the heart's right atrium, so changes in atrial pressure are reflected in the jugular waveform. Distortions in the JVP provide insight into cardiac function without direct assessment of the heart itself, such as resistance diseases (e.g., pulmonary hypertension, tricuspid stenosis~\cite{gibson2005,fang2014book}), mechanical diseases (e.g., tricuspid regurgitation~\cite{fang2014book}), electrical diseases (e.g., atrial fibrillation, heart block, atrioventricular dissociation~\cite{fang2014book}), abnormal external forces (e.g., tamponade, constrictive pericarditis~\cite{fang2014book,welch2015}), and heart failure~\cite{lee1991}.

The main problem with assessing JVP lies in the current method of measurement: invasive catheterisation. Catheterisation requires surgically inserting a central line into either the jugular vein, superior vena cava, or right atrium. This is an invasive procedure requiring surgical expertise. Therefore, although the JVP can provide important clinical insights, JVP examination is not routine and is only performed when there is probable cause for monitoring. Additionally, catheter monitoring is limited to measuring a single location, and does not provide spatial trajectory information. Ultrasound has recently been proposed to measure the JVP through Doppler velocity imaging~\cite{sisini2015,zamboni2016}. However, these methods require constant stable probe contact, expensive ultrasound equipment, trained ultrasound technicians, and is only able to provide single-slice hemodynamic information.

Non-contact cardiovascular monitoring systems that may provide touchless, non-invasive JVP monitoring are photoplethysmographic imaging (PPGI) systems. Building on photoplethysmography theory~\cite{allen2007}, PPGI systems comprise a camera and an illumination source, and leverage the properties of light-tissue interaction to assess blood flow in the skin. These systems have recently gained interest due to their non-contact light-based nature. However, use of PPGI systems have been limited to extracting high-level vital signs, such as heart rate~\cite{poh2010,humphreys2007,sun2011,dehaan2013,xu2014} and respiratory rate~\cite{sun2011,poh2011}. The few studies that have reported spatiotemporal perfusion have been limited to perfusion through the hand in constrained environments~\cite{kamshilin2011,kamshilin2013}. Analysing the spatiotemporal trajectory patterns in the neck can provide insight into whether PPGI can be used to detect the JVP. To this end, we know of no publication that investigated the use of PPGI systems for JVP monitoring. In fact, we know of no study that has explored leveraging PPGI to assess venous blood flow.

Coded hemodynamic imaging (CHI) is a new PPGI system that is optimised for assessing local spatiotemporal blood pulse trajectories~\cite{amelard2015scirep,amelard2016vislet}. CHI is a safe light-based optical system that is able to assess spatiotemporal trajectories by controlling the spatial and/or temporal properties of the illumination source and employing image processing algorithms. Motivated by the clinical promise of non-invasive assessment of JVP, CHI can be used to assess whether the jugular pulse can be observed in the neck, as well as information about its spatial trajectory.

In this study, CHI was used to assess the spatiotemporal pulsing patterns in the neck in 24~participants. Strong pulsatile flow was successfully observed, consistent with ground truth finger photoplethysmography measurements. Two different types of pulsatile flow were observed in all participants: one that corroborated with the ground truth arterial pulse, and one that exhibited strong inverted pulsing characteristics. The hypothesis was that the two pulsatile flows were major arterial and venous blood flow waveforms. To test this hypothesis, the spatial properties of the pulsing were compared to the carotid artery and jugular vein track found through ultrasound, and the waveforms were compared to the JVP. It was concluded that CHI was able to observe major venous flow in all participants. This is the first time to our knowledge that both arterial and venous flow has been reported using a PPGI system. Furthermore, the jugular pulse waveform was observed at many different locations along the venous track, providing indication of the pulse trajectory. These findings may be leveraged in the future as a new way to assess cardiac function in a non-contact manner.

\section*{Results}
\label{sec:results}
\subsection*{Detection of pulsatile flow}
\begin{figure}
\centering
\includegraphics[height=5cm]{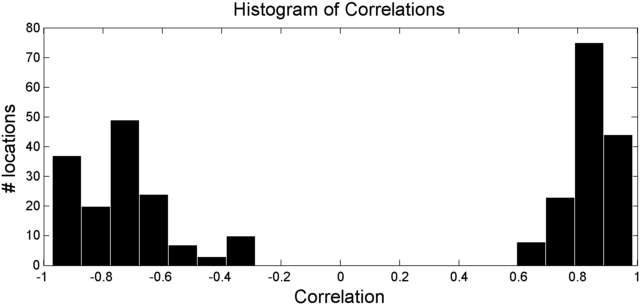}
\caption{Histogram of the strongest signal correlation values relative to ground truth arterial waveform across all participants ($n=24$). The distribution is strongly bimodal, indicating one group of signals that are highly correlated with the arterial waveform, and another group that is strongly negatively correlated with the arterial waveform.}
\label{fig:fig1}
\end{figure}

To test the base hypothesis that CHI was able to detect localised pulsatility in vascularised locations, each $5 \times 5$~mm region was analysed for pulsatile blood flow. Pearson's linear correlation coefficient~($\rho$) was calculated between the ground truth waveform and each region's waveform. The strongest 2.5~cm$^2$ tissue area of positively and negatively correlated pulsatile regions were identified, totalling 5~cm$^2$ tissue area. Figure~\ref{fig:fig1} shows the correlation distribution of the strongest 5~cm$^2$ total tissue, across all participants. The distribution is strongly bimodal, indicating both strong arterial and inverted pulse signals relative to the ground truth PPG. That is, some signals exhibited strong positive correlation to the PPG waveform, whereas other signals exhibited strong negative correlation to the PPG waveform. Strong signals ($|\rho|>0.5$) were found in most participants. The weakest signals were found in participants with high body fat content and age-related skin inelasticity. 

\begin{figure}
\centering
\includegraphics{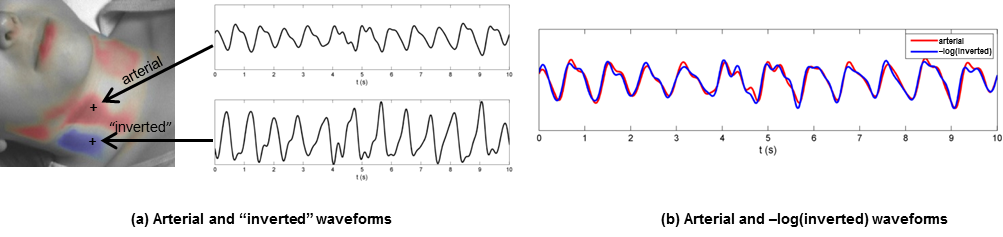}
\caption{Comparison of scale-normalized arterial and inverted pulse waveforms in a typical example. (a) The participant exhibited strong arterial pulsations, characterised by a sharp rise to systole and a dicrotic notch, as well as strong inverted pulsations, characterised by a phase-offset gradual rise and sharp drop. (b) The arterial and inverted waveforms are strongly linked through an inverse relationship ($\rho=0.96$). Best viewed in colour.}
\label{fig:fig3}
\end{figure}

Characteristics commonly found in major arterial waveforms were observed in the positively correlated waveforms, including a sharp increase in absorbance toward a systolic peak, followed by a dicrotic notch, then a diastolic minimum. The ``inverted'' waveforms exhibited different characteristics, specifically a gradual rise, followed by a steep drop in absorbance, and often a small fluctuation following the drop. These rises and drops were temporally aligned. Figure~\ref{fig:fig3} shows a typical example of the two types of pulses observed in a participant. The two pulses were strongly correlated through an inverse relationship ($\rho=0.96$), and each exhibited strong spatially cohesive patterns. Extended Data Figure~1 shows the inverted pulse signal acquired for all participants. The hypothesis was that CHI was detecting both arterial and venous blood pulses. The following subsections tested this hypothesis.

\begin{figure}
\centering
\includegraphics{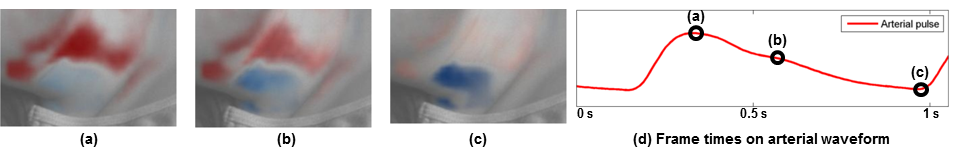}
\caption{Three frames from a typical segment showing the timing of the blood volume pulse from Extended Data Video~1. (a) Upon ventricular ejection, the pulse travels through the carotid arterial track (red), reaching systole; there is no inverted pulsatile flow. (b) During the arterial descent towards diastole, the pulse continues to travel through the carotid track, and the start of the inverted pulse can be observed (blue). (c) No carotid pulse is observed during diastole, and the inverted pulse experiences its maximum absorbance. Best viewed in colour.}
\label{fig:fig4}
\end{figure}

\subsection*{Primary anatomical locations of strong pulsing}
Figure~\ref{fig:fig4} shows selected frames from a video of the flow profiles, Extended Data Video~1. The two types of flow were easily identified visually due to the phase offset nature of their peaks. Specifically, the arterial track filled with blood simultaneously as the ground truth arterial waveform reached systole. When the arterial waveform reached diastole, the inverted pulse reached its peak. The two tracks exhibited consistent alternating pulsing patterns.

\begin{figure}
\centering
\includegraphics{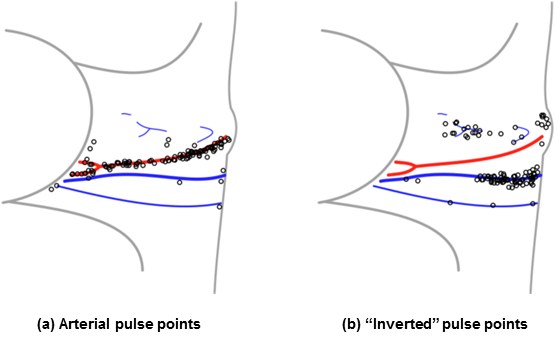}
\caption{Locations exhibiting the strongest pulsing across all participants ($n=24$) relative to the position of the carotid artery and jugular vein. Locations were determined by finding and marking the carotid and jugular track using ultrasound following video data capture. The pulsing locations were normalized and drawn relative to the individual's marked anatomy. The data are presented for the right side only (marked with `o'), which is used clinically as the most direct conduit to the heart. The arterial pulse points were consistent with the anatomical location of the carotid artery (a), and the inverted waveform pulse points were consistent with the distal location of the jugular vein (b). Best viewed in colour.}
\label{fig:fig5}
\end{figure}

Figure~\ref{fig:fig5} shows the locations on the neck where the five strongest positive and inverted pulsations occurred for each participant. Following data collection, the carotid and jugular tracks were marked with the guidance of ultrasound. Comparing the marked locations with the major pulsing locations, the arterial pulse locations followed the carotid track. Inverted pulsatile flow was consistently situated on the distal side of the carotid artery. This was consistent with cross-sectional ultrasound imaging, which located the jugular vein on the distal side of the common carotid artery in all participants.

During data analysis, Doppler ultrasound confirmed that the jugular vein was pulsatile in all 24 participants, and cross-sectional ultrasound analysis visually confirmed the phase offset nature of the carotid and jugular pulsing.

\subsection*{Correlation to jugular venous pulse waveform}
\begin{figure}
\centering
\includegraphics{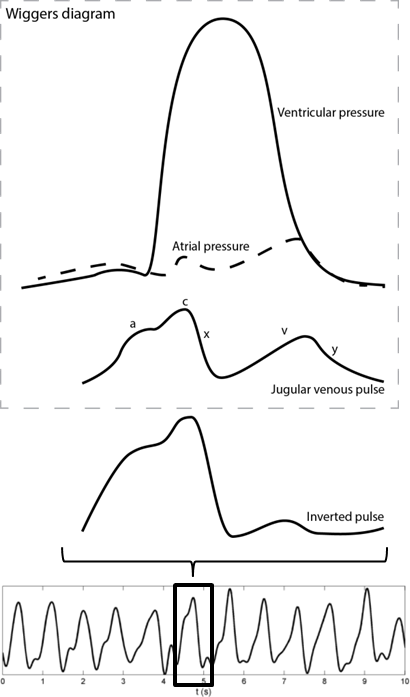}
\caption{Comparison of a typical ``inverted'' pulse to the jugular venous pulse waveform in the Wiggers diagram (adapted from~\cite{opie2014book}). The inverted pulse was consistent with the JVP waveform. The JVP waveform is biphasic whose waveform inflections are governed by differential cardiac pressures (see text). The variability between individual pulses is likely due to the effect of respiration on the intrathoracic pressure.}
\label{fig:fig6}
\end{figure}

The inverted blood pulse waveform shape was consistent with the jugular venous pulse (JVP) waveform~\cite{walker1990book}. Figure~\ref{fig:fig6} shows the time-aligned Wiggers diagram section with the labeled JVP and a typical inverted pulse observed during trials. Visually, the JVP and inverted pulse waveforms show strong intercorrelation. The JVP waveform is biphasic and is characterised by: an increase in pressure pre-systole due to right atrial contraction (a); an increase in pressure due to ventricle contraction (c); a decrease in pressure during systole due to atrial relaxation following tricuspid valve closure (x); an increase in pressure in late systole due to right atrial filling from venous return (v); and a decrease in pressure during diastole due to right ventricular filling from the opening of the tricuspid valve (y). The inverted pulsing results were consistent with the JVP waveform: gradual increase in blood volume late systole and diastole (a); an increase in pressure during ventricular systole (c); a sharp decrease in blood volume was observed slightly prior to the carotid upstroke (x); and a small transient rise in blood volume during late systole (v,y). Since the JVP is governed by differential pressures generated by heart mechanics, observing the venous pulsation patterns can provide insight into not only vascular function, but aspects of heart function as well, without catheterisation.


\section*{Discussion}
\label{sec:discussion}
These trials support the hypothesis that CHI was able to observe the spatial trajectory of both major arterial and venous blood pulse waveforms in the neck. The amplitude changes of the two signals exhibited different differential properties. In particular, the arterial signal was characterised by a sharp rise during systole followed by a fall during late systole and diastole, whereas the venous signal was characterised by a gradual rise during atrial filling followed by a sharp drop during atrial and ventricular contraction. These amplitude fluctuations reflect changes in blood volume, which change the amount of absorbed light~\cite{allen2007}.

\begin{figure}
\centering
\includegraphics{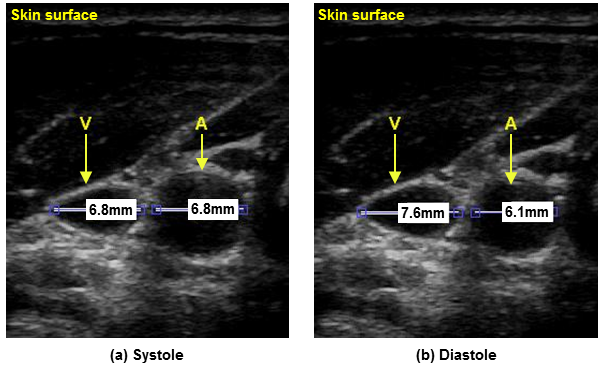}
\caption{Inverse relationship between the carotid artery and jugular vein as seen with B-mode ultrasound in a typical participant. The jugular vein (below V) and carotid artery (below A) are located below the skin surface (top) with vessel diameters marked in millimetres. Between systole (a) and diastole (b), the carotid relaxes (10\% diameter reduction), and the jugular expands (12\% diameter expansion).}
\label{fig:fig10}
\end{figure}

The strong negative correlation between the arterial and venous blood pulse waveforms can be attributed to the differential pressure profiles resulting from normal cardiac cycles. The observed results are consistent with the Wiggers diagram. In particular, cardiac contraction ejects blood through the arterial track, and ends with aortic closure. During arterial systole, atrial relaxation causes a decrease in cardiac pressure, resulting in increased venous return into the right atrium, which reduces the volume in the neck veins. Atrial filling pressure gradually increases simultaneously with late systole, resulting in decreased venous return (increased jugular venous volume). This inverted amplitude and phase-offset vessel wall motion was visually confirmed using B-mode ultrasound, where a cross-sectional movie showed phase-offset vessel wall expansion and contraction between the carotid artery and jugular vein (see Figure~\ref{fig:fig10}). Doppler ultrasound was also used to confirm the pulsatile nature of the jugular vein in the participants consistent with the JVP waveform.

Due to the close proximity of the major neck vessels to the heart, the venous blood pulse waveform can be used to assess heart function. Leveraging the strong correlation between the venous blood pulse waveform and JVP, CHI can be used to assess heart function that is reflected in the venous waveform in a non-contact manner. Since the jugular vein is a major venous extension of the right atrium, abnormalities in the waveform can indicate heart function problems. For example, the increased pressure from atrial contraction due to tricuspid stenosis produces a larger a-wave~\cite{fang2014book,gibson1955}; a lack of atrial contraction/relaxation from atrial fibrillation inhibits the a-wave and x-wave~\cite{fang2014book}; and blood flow back through the atrium due to tricuspid regurgitation results in a fused c-v wave~\cite{fang2014book}.


These findings show promise for shifting from catheter insertion techniques for measuring the JVP waveform toward a non-contact, non-invasive light-based imaging solution, making routine examinations feasible. As opposed to single-point catheter measurements, CHI can provide spatial properties of the JVP. This type of solution promotes the assessment of the jugular pulse in non-surgical settings, and can be used to analyse the spatial trajectory properties of the jugular pulse. Future work can leverage these findings to investigate cardiac dysfunction for rapid visual patient assessment.

\section*{Methods}
\subsection*{Study protocol}
Data were collected across 24 participants (age $(\mu \pm \sigma)=28.7\pm12.4$). Extended Data Figure~2 graphically shows the setup of the study. Demographic information (age, height, weight, body fat \%) was obtained at the beginning of the study. The participants were asked to assume a supine position for the duration of the study. The ground truth blood pulse waveform was collected using the Easy Pulse finger photoplethysmography (PPG) cuff simultaneously with the video data. B-mode ultrasound (Vivid i, General Electric Healthcare, Horten, Norway) was used to confirm the location of the jugular vein relative to the carotid artery after the video data were collected. Cross-sectional videos were acquired to show the relative position of the carotid artery and jugular vein, and to visualise the pulsatility of the vessel walls. Doppler measurements were acquired of the jugular vein to investigate jugular flow velocity. The carotid and jugular paths were marked after video collection. Informed consent was obtained from all participants, and by those participants whose photos were used in this paper. The study was approved by a University of Waterloo Research Ethics committee.

\begin{figure}
\centering
\includegraphics[width=0.6\textwidth]{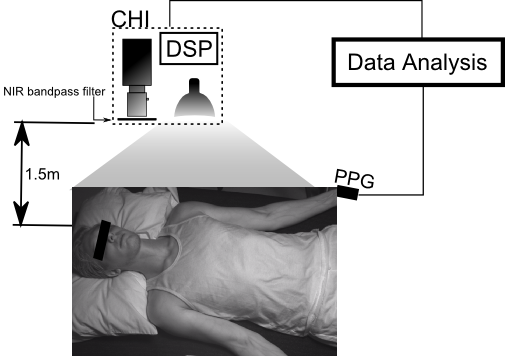}
\caption{Study setup. Participants were supine for the duration of the study. The imaging system (CHI) was positioned above and slightly to the right of the participant, at a distance of 1.5~m. Illumination was provided by a spatially uniform 250~W tungsten-halogen illumination source. Imaging data were processed on a digital signal processing (DSP) unit. The participant wore a finger cuff which provided the ground truth arterial waveform for the analysis.}
\label{fig:fig8}
\end{figure}

\begin{figure}
\centering
\includegraphics[width=0.8\textwidth]{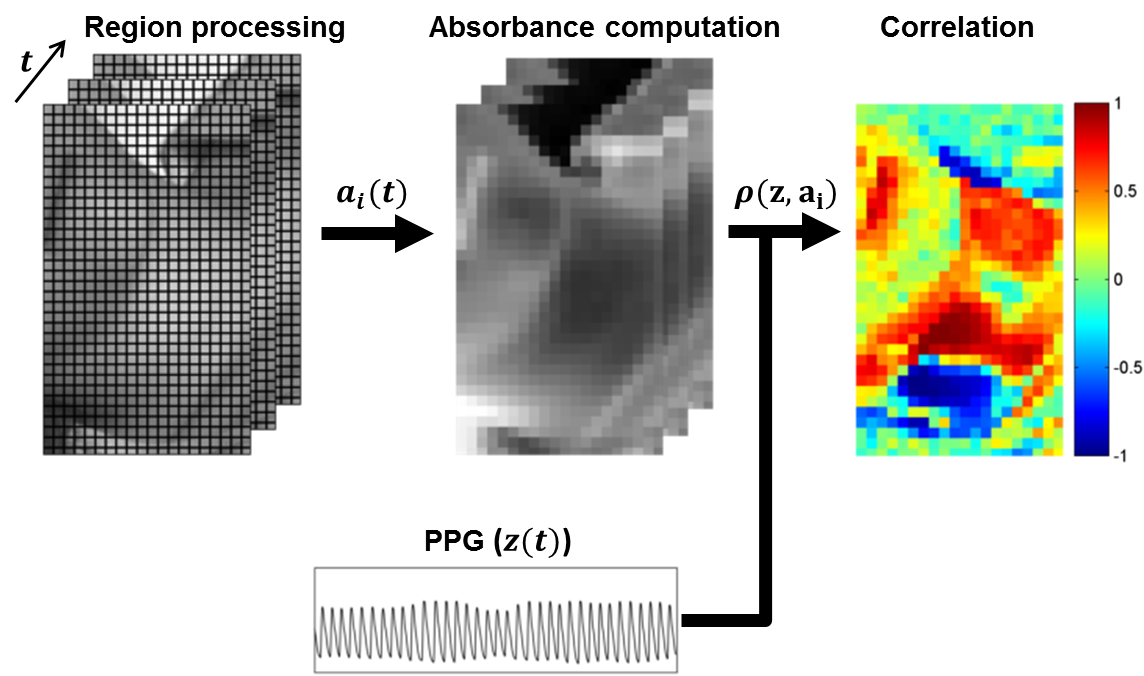}
\caption{Data processing pipeline. Each frame was analysed in $0.25\times0.25$~mm regions. Each region's temporal fluctuations were converted to absorbance using~(\ref{eq:ai}). Pearson's linear correlation coefficient was computed for each waveform using the ground truth PPG waveform, yielding a spatial correlation map showing locations exhibiting strong forward and inverted pulsing. Best viewed in colour.}
\label{fig:fig9}
\end{figure}

\subsection*{Coded hemodynamic imaging (CHI)}
A novel coded hemodynamic imaging (CHI) system was used to collect the data. CHI consisted of a near infrared sensitive camera (PointGrey GS3-U3-41C6NIR-C), an 850--1000~nm optical bandpass filter, and a 250~W tungsten-halogen illumination source. The illumination was spatially coded with a diffuse uniform pattern using a 16'' glass fabric front diffuser. Both the illumination source and CHI were situated 1.5~m above the participant. In one case, the participant's neck was not visible from overhead, so a bedside view was used. Videos were collected at 60~fps, using 16~ms exposure time and f4.0 aperture. The data were processed using a digital signal processing (DSP) unit.

Extended Data Figure~3 shows the signal processing pipeline for the study. Each frame was blockwise averaged using $5\times5$~mm regions. The temporal fluctuations of region $i$ yielded a reflected illumination signal $x_i(t)$. Reflectance was converted to absorbance using Beer-Lambert law, modeling the geometrical photon path as the scattered reflected path~\cite{pellicer2011}:
\begin{equation}
\label{eq:ai}
  a_i(t) = -\log \left( \frac{x_i(t)}{x^0_i(t)} \right)
\end{equation}
where $x^0_i(t)$ denotes the illumination incident on the tissue at time $t$. A detrending method~\cite{tarvainen2002} was used to eliminate gradual deviations in ambient conditions.

\subsection*{Data analysis}
Using the finger PPG signal as the ground truth arterial blood pulse waveform, the Pearson's linear correlation coefficient was computed between each region's absorbance signal $a_i(t)$ and the PPG signal $z(t)$ to determine the signal strength and directionality:
\begin{equation}
\label{eq:rho}
  \rho(z,a_i) = \frac{\sigma_{z,a_i}}{\sigma_{z}\sigma_{a_i}} \in [-1,1]
\end{equation}
where $\sigma_{z},\sigma_{a_i}$ are the standard deviation of the PPG sensor and region signals respectively, and $\sigma_{z,a_i}$ is the covariance of the two signals. $\rho>0.5$ indicates strong pulsing consistent with the PPG arterial pulsing signal, and $\rho<-0.5$ indicates strong pulsing that is inversely proportional to the PPG signal. For visualisation purposes, the colour maps were smoothed using a Gaussian kernel ($\sigma=2.5$~mm).

In some cases, the carotid arterial waveform's shape differed substantially from the finger PPG waveform (e.g., the finger did not exhibit the dicrotic notch). Thus, the waveform that exhibited the strongest signal-to-noise ratio (SNR) with respect to the PPG waveform was used as a template. SNR was calculated in the frequency domain across all regions as:
\begin{equation}
  SNR=10 \log_{10} \left( \frac{\sum_f (Z(f))^2}{\sum_f (Z(f)-A_i(f))^2} \right)
\end{equation}
where $Z,A_i$ are the zero-DC normalized spectral magnitudes of the PPG and $i^\text{th}$ region signals, respectively, and $f$ represents frequency.

\section*{Acknowledgements}
\label{sec:ack}
The authors thank Ikdip Brar for her help with data collection. The study was funded by the Natural Sciences and Engineering Research Council (NSERC) of Canada (RGPIN-6473, CGSD3-441805-2013), AGE-WELL NCE Inc. (AW-HQP2015-23), and in part by the Canada Research Chairs program.

\section*{Author contributions}
R.A., A.W. designed the system. J.L. built the system. R.A., A.W. designed the algorithms. R.A., R.H., D.G., K.P., A.W. were involved in the data acquisition protocol. R.A., R.H., D.G., K.P., A.W. conceived the hypothesis. R.A. analysed the data. R.A. drew figures 4 and 5. All authors reviewed the manuscript.

\section*{Additional Information}
\textbf{Competing financial interest:} All authors in this study have no competing financial interests.

%
%

 \bibliographystyle{naturemag}
\bibliography{bib}

\end{document}